\documentclass[superscriptaddress]{revtex4-2}

\usepackage{amsmath} 
\usepackage{amssymb}  
\usepackage{amsfonts} 
\usepackage{dsfont}
\usepackage{bm}  
\usepackage{bbm}   
\usepackage{bbold}   
\usepackage{braket} 
\usepackage{color} 
\usepackage{comment}  
\usepackage{dcolumn}  
\usepackage{enumerate}  
\usepackage{epsfig}  
\usepackage{gensymb}  
\usepackage{graphicx}  
\usepackage{indentfirst}  \usepackage{lmodern}  
\usepackage{mathrsfs}  
\usepackage{mathtools}  
\usepackage{psfrag}  
\usepackage{pst-all} 
\usepackage{soul}  
\usepackage{xcolor}
\usepackage{float} 
\usepackage[colorlinks,linkcolor=blue,citecolor=blue,urlcolor=blue,hyperindex,driverfallback=dvipdfm]{hyperref}  \usepackage[T1]{fontenc} 

\usepackage{dirtytalk}
\usepackage{CJKutf8}

\usepackage{float} 
\usepackage{placeins}
\usepackage{bm}
\usepackage{enumitem}

 \usepackage[normalem]{ulem} 
 \makeatletter
\newcommand*{\addFileDependency}[1]{
  \typeout{(#1)}
  \@addtofilelist{#1}
  \IfFileExists{#1}{}{\typeout{No file #1.}}
}
\makeatother

\usepackage[strict]{changepage}
\usepackage{calc}
\usepackage{dirtytalk}
\usepackage{pifont}

\usepackage{xr-hyper}

\usepackage{graphicx}
\usepackage{cancel}
\usepackage{longtable}
\usepackage{array}
\newcolumntype{C}[1]{>{\centering\arraybackslash}p{#1}}

\usepackage{tensor}

\usepackage{ulem}
\usepackage{titlesec}
 \titleformat{\paragraph}[hang]{\bfseries}{}{0pt}{\uline}
\titlespacing*{\paragraph}
{0pt}{3.25ex plus 1ex minus .2ex}{1.5ex plus .2ex}

\usepackage{dirtytalk}
\usepackage{cleveref}
\usepackage[]{siunitx}
\usepackage{textcomp}
\usepackage{upgreek}
\begin{document}

\title{ Quantum Dial for High-Harmonic Generation }
\author{Lu Wang*}
\affiliation{Department of Physics, University of Ottawa, Ottawa, Ontario K1N 6N5, Canada}
\email{lu.wangTHz(at)outlook.com}
\author{Andrew M Parks}
\address{Wyant College of Optical Sciences, University of Arizona, Tucson, Arizona,  USA}
\author{Adam Thorpe}
\affiliation{Department of Physics, University of Ottawa, Ottawa, Ontario K1N 6N5, Canada}
\author{Graeme Bart}
\affiliation{Department of Physics, University of Ottawa, Ottawa, Ontario K1N 6N5, Canada}
\author{Giulio Vampa}
\address{Joint Attosecond Science Laboratory, National Research Council of Canada and University of Ottawa, 100 Sussex Drive, Ottawa,
Ontario K1A 0R6, Canada}
\author{Thomas Brabec}
\affiliation{Department of Physics, University of Ottawa, Ottawa, Ontario K1N 6N5, Canada}
\begin{abstract}
High-harmonic generation (HHG) is an extreme nonlinear optical process in which intense laser fields generate radiation at integer multiples of the driving frequency. Recent advances in bright squeezed vacuum (BSV) sources have opened the possibility of driving and controlling HHG with nonclassical light. Here, we introduce a hybrid scheme in which a strong classical field drives the nonlinear dynamics, while a much weaker BSV pulse provides quantum-optical control. Remarkably, a BSV field with an amplitude approximately $50$ times smaller than that of the classical driver acts as an effective "optical dial", enabling substantial control over the harmonic spectrum and its angular dependence. We further show that combining a linearly polarized classical field at $\omega_0$ with a linearly polarized BSV field centered at $0.5\omega_0$ produces pronounced valley-contrasting responses. These results establish weak quantum fields as powerful control parameters for extreme nonlinear dynamics and suggest new opportunities for valley-based information processing beyond conventional charge-based electronics.
\end{abstract}

\maketitle

When an intense laser field interacts with matter, it generates emissions at integer multiples of the driving laser frequency—a process known as high-harmonic generation (HHG). HHG has led to many successful applications, such as attosecond pulse generation in the extreme ultraviolet and soft x-ray regions \cite{drescher2001x,sansone2011high,li2020attosecond,agostini2004physics,paul2001observation} and time-resolved spectroscopy \cite{peng2019attosecond,wagner2006monitoring,ozawa2008high}. {Semiclassical descriptions of HHG are typically based on the time-dependent Schrödinger equation or the Lewenstein strong-field approximation for atomic and molecular gases \mbox{\cite{corkum1993plasma,lewenstein1994theory}}, where the electron dynamics are treated either fully quantum mechanically or within a single-active-electron approximation. For solids, HHG is commonly described using the semiconductor Bloch equations (SBE), which account for coupled interband and intraband dynamics \cite{vampa2014theoretical,vampa2015linking}}. Additionally, time-dependent density functional theory incorporates many-body electron–electron interactions, providing a more realistic description of the electronic dynamics at substantially higher computational cost \cite{tancogne2017impact,goings2018real}. These models have so far qualitatively predicted harmonic features such as the plateau, the cutoff energy, polarization dependence, and the pulse duration, in alignment with experimental measurements \cite{you2017anisotropic,dromey2006high,shiner2009wavelength,tong2000probing,Ghimire2011NatPhys,Ghimire2019NatPhys}. 

Very recently, owing to the advancement in obtaining strong femtosecond quantum pulses  (nJ-\textmu J) \cite{finger2015raman,gorlach2023high,sh2012superbunched,rasputnyi2024high,kern2025single}, HHG has been demonstrated with quantum light \cite{rasputnyi2024high,finger2015raman,gorlach2023high,even2023photon}. In particular, there has been considerable interest in the bright squeezed vacuum (BSV), which is a macroscopic quantum state of light generated by a strongly pumped unseeded optical parametric amplifier \cite{rasputnyi2024high,fabre2020modes,agafonov2010two,sharapova2020properties}. BSV exhibits remarkable quantum properties such as pronounced photon-number correlations, quadrature squeezing, and polarization entanglement \cite{agafonov2010two,tzur2025measuring,lemieux2025photon}. Existing studies suggest that, compared to a classical driving field, HHG driven by BSV possesses unique characteristics such as quantum correlations between electrons and photons, photon bunching, and extended cutoff \cite{gothelf2025high,lemieux2025photon,tzur2025measuring,gorlach2023high}. It has been shown that through its intrinsic photon-number fluctuations and correlations, quantum light can strongly enhance nonlinear responses at low pulse energies, enabling regimes inaccessible to classical light while mitigating material damage \cite{rasputnyi2024high,jiang2026nonlinear}. Very recent research has shown that the combination of classical and quantum fields can distort electron trajectories and result in bunched harmonic photons \cite{lemieux2025photon}.

Most existing studies mentioned have focused on HHG driven solely by intense classical or quantum fields. Here, we consider a strong classical pulse accompanied by a weak quantum pulse \cite{lemieux2025photon}. Specifically, the classical field has amplitude $\mathrm{E}_0=3\times10^9,\mathrm{V/m}$ ($1.2\,\mathrm{TW/cm^2}$), while the quantum perturbation has $\mathrm{E}_s\sim5\times10^7\,\mathrm{V/m}$, corresponding to $\sim$1 nJ. Despite its low energy, the quantum light serves as an effective optical tuning knob, enabling dramatic control of phenomena that only arise from quantum perturbations. To investigate the fundamental physics of the BSV perturbation on harmonic emission, we focus on two-dimensional materials. In bulk materials, macroscopic nonlinear propagation effects, such as interference between emissions from different regions within the nonlinear material, can significantly alter the harmonic signal \cite{wang2024tabletop}. These bulk effects may conceal the underlying microscopic mechanisms. In 2D materials, such complications are reduced, allowing the microscopic dynamics of high-harmonic generation to be more clearly isolated. In particular, we have chosen the transition metal dichalcogenides, which demonstrate remarkably strong optical nonlinearities that allow access to significant optical responses with moderate driving field strength \cite{le2018high,liu2017high,mak2010atomically}. 

Combining optical control using BSV and two-dimensional materials offers a promising route toward compact, integrated photonic devices. Our results show that BSV strongly modifies the amplitude and phase of the optical responses throughout the Brillouin zone. We find that the combination of a $\omega_0$ classical field and a $0.5\omega_0$ BSV perturbation can also induce pronounced valley-contrasting responses. Moreover, the harmonic emissions can be controlled by varying the relative angle between the laser and BSV polarizations. Furthermore, because the method allows accessing fairly high-order harmonics with very weak perturbations, it can be integrated with high-repetition-rate lasers, potentially enabling table-top diagnostic tools for nonlinear spectroscopy of materials while mitigating issues such as laser-induced damage or heating.

Our theoretical model treats the material responses as a 2-band density matrix embedded in an open quantum system, driven by a classical laser field. The many-body nature of the quantum environment (e.g., the BSV) is accounted for as a perturbation \cite{boroumand2025strong}. With the convenience and power of the theoretical framework we have developed, quantum light can be treated very effectively via a simple scalar function --- the response function. This significantly reduces the computational and mathematical complexity associated with quantum light-related physics.

\section{Results}
\subsection{Control of HHG via quantum light}
\begin{figure}[H]
\centering
\includegraphics[width=1\linewidth]{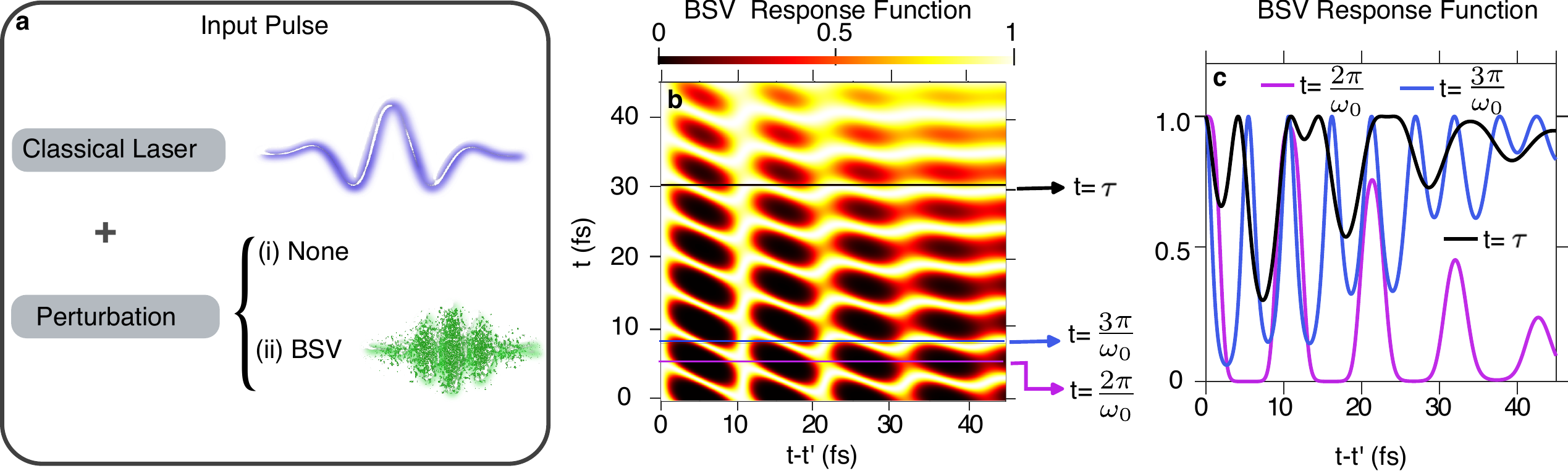}
\caption{Panel \textbf{a} illustrates the concept of the proposed scheme, i.e controlling nonlinear interactions by a weak quantum perturbation (bright squeezed vacuum). Panel \textbf{b} and \textbf{c} show the response functions $\mathcal{R}(t',t)$, with panel \textbf{c} presenting its values at selected times from panel \textbf{b}. The data is calculated with BSV center frequency $\omega_s=\omega_0$.}\label{fig:response_func}
\end{figure}

Our motivation is to significantly modify the nonlinear interaction via a very weak quantum perturbation such as the BSV, illustrated in Fig.\ref{fig:response_func}\textbf{a}. We focus on two cases where the unperturbed case, driven solely by the classical field, is denoted as "(i) None" and serves as the reference; the classical pump accompanied by a weak BSV is referred to as "(ii) BSV". In particular, the classical laser has a fixed center 
wavelength $\lambda_0=3.2\, $\unit{\micro\meter} throughout this work, which corresponds to the center angular frequency $\omega_0= 2\pi \times 10^{14}$\,Hz  ($\hbar\omega_0 =0.39$\, eV) with $c$ the speed of light in vacuum. We use a linearly polarized classical field defined as
${E}_0\exp{\left(-t^2/\tau^2\right)}\cos{(\omega_0t)}$, where $\tau=30\,\text{fs}$, and ${E}_0= 3\times10^9\,\text{V}/\text{m}$ ($1.2 \text{TW}/\text{cm}^2$). The BSV contains energy  $\sim 1$ nJ, corresponding to a photon number $ \sim 10^ {10} $ and field strength $E_s \sim 5\times10^7$ V/m (see Supplementary Material, Section II). Because the quantum field amplitude is $\sim$ 50 less than the classical field, an equally weak classical perturbation would produce a negligible change in the nonlinear response. Consequently, any observable contrast between cases (i) and (ii) can be attributed to the quantum nature of the perturbation.

Commonly, in solid-state materials, the interactions are modeled by the Semiconductor Bloch wave equation \cite{vampa2014theoretical,thorpe2023high,boroumand2025strong,wang2024tabletop}, within the independent particle approximation. Recently, it was shown that a wide class of many-body perturbations can be approximately accounted for via a bosonic environment \cite{boroumand2025strong}. This model is generalized here to describe the quantum field, which is represented by an ensemble of boson modes.

With our model, the quantum perturbation is condensed into a scalar function --- the response function $\mathcal{R}$, which is purely real for very large photon numbers [see Eqs(\ref{main:eq:Rs}-\ref{main:eq:gq}) and Supplementary Material Sections V and VI]. The emitted current is proportional to the term 
\begin{equation}\label{main:eq:response_illustrate}
  \int_{-\infty}^t  \exp{\left[2iS(t,t^\prime)\right]}\mathcal{R}(t,t^\prime)\Omega(t^\prime)dt^\prime
\end{equation}
where $\Omega$ is the Rabi frequency.  In particular, the response function  $\mathcal{R}$ is time-dependent and is also of the exponential form $\exp{( )}$. As shown by Eq.(\ref{main:eq:response_illustrate}), the exponent of  $\mathcal{R}$ enters the electron dynamics in the same way as the action term $S(t,t^\prime)$ described via the Lewenstein model \cite{lewenstein1994theory,boroumand2025strong}. In the case of (i), where there is no perturbation, the response function simplifies to  $\mathcal{R}=1$. As a scalar function of two time variables, $\mathcal{R}(t,t^\prime)$ is represented as a two-dimensional distribution in Fig. \ref{fig:response_func}\textbf{b}. It is important to notice that a BSV with center frequency $\omega_0$ corresponds to a response function that contains two dominant oscillation frequencies $\omega_0$ and $2\omega_0$ (see Supplementary Material Section VI.B.). By denoting $T_0=2\pi/\omega_0$, we choose a few representative values $t=T_0$, $t=3T_0/2$, and $t=\tau$ to present in Fig. \ref{fig:response_func}\textbf{c}. When $t$ equals $t^\prime$, $\mathcal{R}(t=t^\prime)=1$. It can also be seen that the response function is non-Markovian (not instantaneous).

\begin{figure}[H]
\centering
\includegraphics[width=1\linewidth]{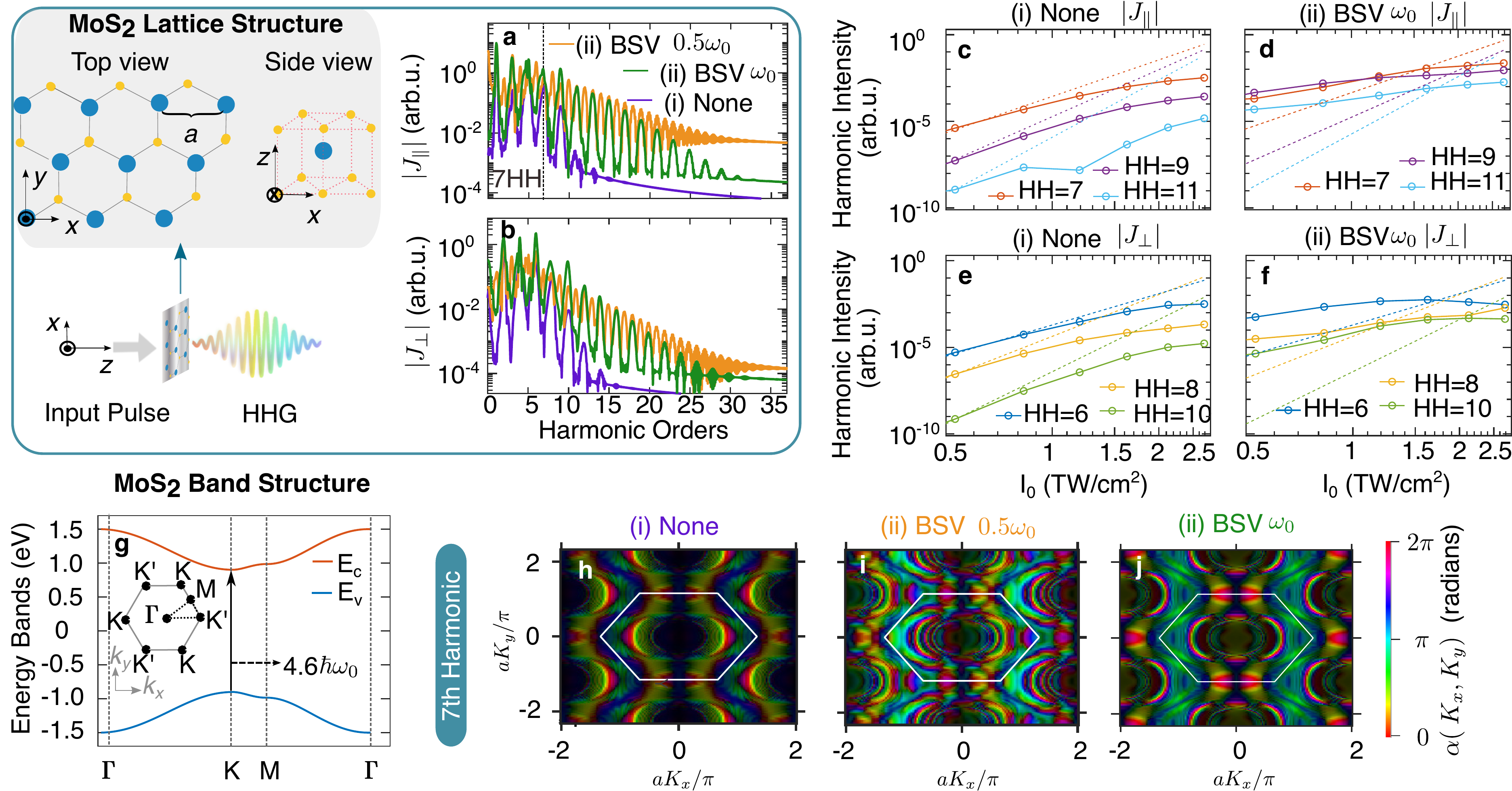}
\caption{The real-space lattice structure of MoS$_2$ is shown in the top-left corner, and its corresponding band structure is presented in panel \textbf{g}. When both the classical field and the BSV are polarized along $x$ (perpendicular to the mirror symmetry axis of the MoS$_2$), the resulting parallel ($x$, $J_\parallel$) and perpendicular ($y$, $J_\perp$) harmonic emissions are presented in panels \textbf{a} and \textbf{b}, respectively. Different colors represent three cases: "(i) None", classical pulse only; "(ii) BSV $\omega_s=0.5\omega_0$", where the BSV central frequency is half that of the classical pulse; and "(ii) BSV $\omega_s=\omega_0$", where the BSV and classical pulse have the same central frequency. Panels \textbf{c–f} show the harmonic yields as a function of the classical pulse intensity $I_0$. The dashed curves indicate the perturbative scaling $I_0^n$ where $n$ is the harmonic order. The optical responses over the entire Brillouin zone of the 7th-order harmonic (see dashed vertical line in panel \textbf{a}) are presented in panels \textbf{h}-\textbf{j}. The white hexagon marks the boundary of the first Brillouin zone.}\label{fig:setup_compare}
\end{figure}%

To proceed with the light-matter interaction analyses, we focus on the 2D material--transition metal dichalcogenides. This type of material possesses a hexagonal lattice structure, and the chemical formula MX$_2$, where M (such as Mo, W) is a transition metal and X (such as  S, Se) is a chalcogen (see Fig.\ref{fig:setup_compare} top-left corner). We choose MoS$_2$ due to its relatively weak spin-orbital coupling, making it a suitable candidate for an initial study without the complication of additional effects \cite{zhu2011giant,xiao2007valley,fragkos2025floquet}. The monolayer MoS$_2$ consists of one layer of Mo atoms sandwiched by two layers of S atoms. Though the bulk MoS$_2$ has an inversion center located in the middle of two unit cells between two layers, the MoS$_2$ monolayer does not possess inversion symmetry \cite{xiao2012coupled}. The  MoS$_2$ band structure is calculated via the tight-binding model \cite{ribeiro2011stability,rukelj2016optical,kobayashi2021polarization,nebgen2026isostructural} (details can be found in Supplementary Material Section I). The band structure along $\Gamma$-K-M-$\Gamma$ is shown in Fig.\ref{fig:setup_compare}\textbf{g}, where the conduction and valence bands are denoted by E$_\text{c}$ and E$_\text{v}$, respectively. 

Without loss of generality, both the classical field and the BSV are polarized along the $x$ axis, perpendicular to the mirror-symmetry axis of MoS$_2$. The resulting parallel ($x$, $J_\parallel$) and perpendicular ($y$, $J_\perp$) harmonic emissions are presented in Fig.\ref{fig:setup_compare}\textbf{a} and Fig.\ref{fig:setup_compare}\textbf{b}, respectively. Here, different colors represent three cases: "(i) None", classical pulse only; "(ii) BSV $\omega_s=0.5\omega_0$", where the BSV central frequency $\omega_s$ is half that of the classical pulse $\omega_0$; and "(ii) BSV $\omega_s=\omega_0$", where the BSV and classical pulse have the same central frequency. We can see that the BSV perturbation substantially extends the harmonic spectrum to higher orders \cite{gorlach2023high}. Besides, if the BSV central frequency matches that of the pump $\omega_s=\omega_0$, the resulting spectrum preserves the symmetry of (i) None. When $\omega_s=0.5\omega_0$, additional sidebands appear. Though the response function can contain real and imaginary parts,  for photon numbers $\sim 10^{10}$, the imaginary part is negligible. Our results show that the imaginary part is responsible for the $\pm \omega_s$ transitions, whereas the real part is responsible for the $\pm 2\omega_s$ transitions (see Supplementary Material Section VI.B).

Figure \ref{fig:setup_compare}\textbf{c–f} show the harmonic yields as a function of the classical pulse intensity $I_0$. The dashed curves indicate the perturbative scaling $I_0^n$ where $n$ is the harmonic order. It can be seen that the solid curves deviate from the dashed curves, indicating a non-perturbative regime \cite{liu2017high}. We observe a very weak dependence on the harmonic intensity to $I_0$ when perturbed by a BSV. The selected two-band system has a band gap of 1.8-3\,eV, corresponding to 5th to 8th harmonic orders. Because our model captures the non-perturbative dynamics, we therefore expect enhanced spectral amplitudes within this harmonic range, which agrees with the plateau-like structures seen in Figs. \ref{fig:setup_compare}\textbf{a} and \ref{fig:setup_compare}\textbf{b}  (see Supplementary Material, Sec. VI.A).

To further characterize the influences of the BSV, we show the optical responses over the entire Brillouin zone of the 7th-order harmonic (see dashed vertical line in Fig.\ref{fig:setup_compare}\textbf{a}) in  Fig.\ref{fig:setup_compare}\textbf{h}-\textbf{j}, where the white hexagon marks the boundary of the first Brillouin zone. The complex spectral distribution can be written as $J_x(7\omega_0,K_x,K_y)=|J_x(7\omega_0K_x,K_y)|\exp{[i\alpha(7\omega_0,K_x,K_y)]}$. In these plots, the amplitude $|J_x(7\omega_0,K_x,K_y)|$ is represented by brightness, while the phase $\alpha(7\omega_0,K_x,K_y)$ is encoded by color. The black color suggests the optical responses of the corresponding region are very weak. Note that, for case "(i) None", because MoS$_2$ is symmetric along $x$, the optical responses at $K_x$ and $-K_x$ always have equal amplitudes. For odd harmonics, the responses from each $\pm K_x$ pair have the same phase and therefore add constructively (as in Fig.\ref{fig:setup_compare}\textbf{h-j}). In contrast, for even harmonics, they have opposite phases and cancel when integrated over the Brillouin zone. By summing over the entire momentum space $K_x,\,K_y$, one obtains the nonlinear spectrum in Fig.\ref{fig:setup_compare}\textbf{a}-\textbf{b}. Details can be found in Methods Section A. Eqs.(\ref{main:eq:Jtw},\ref{main:eq:intkxky}). These results suggest that BSV strongly modifies both the amplitude and phase of the response throughout the Brillouin zone.

\subsection{BSV Induced Valleytronic Effects}

\begin{figure}[H]
\centering
\includegraphics[width=0.65\linewidth]{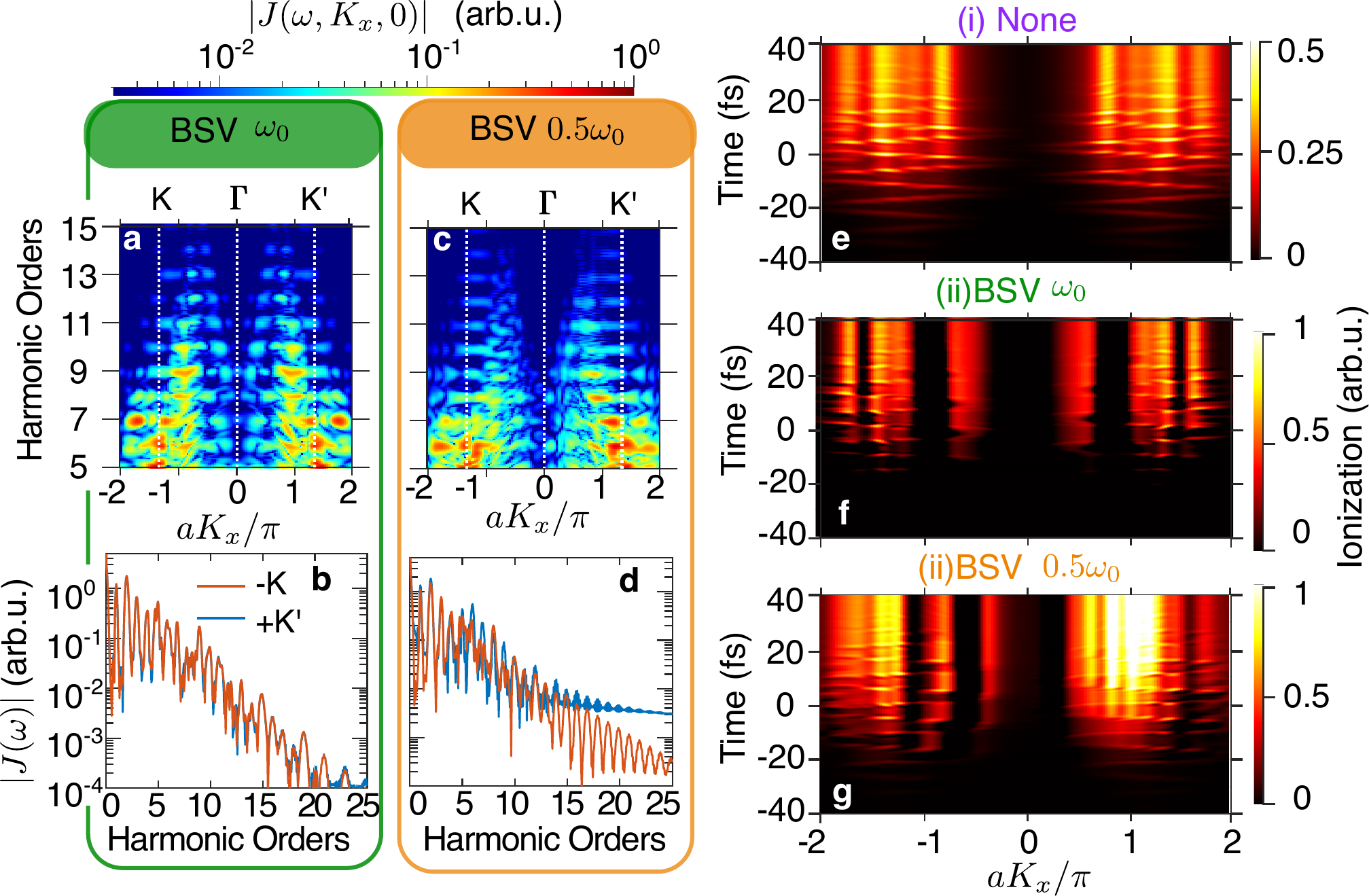}
\caption{Valley-resolved responses along the K$-\Gamma-$K' direction, i.e. $K_y=0$. Panels \textbf{a} and \textbf{c} show the emitted harmonics as functions of $K_x$ for BSV centered at $\omega_0$ and $0.5\omega_0$, respectively. Panels \textbf{b} and \textbf{d} show the spectra integrated from panels \textbf{a} and \textbf{c}, respectively. The spectrum integrated over $K_x>0$ (containing the K') is indicated by the blue curve, whereas that integrated over $K_x<0$ (containing the K) is indicated by the red curve. Panels \textbf{e-g} show the ionization versus $K_x$ for (i)None, (ii) BSV $\omega_0$ and (ii) BSV $0.5\omega_0$, respectively.}\label{fig:valley}
\end{figure}

Here, we show that a linearly polarized classical pump at $\omega_0$, combined with a weak linearly polarized BSV perturbation centered at $0.5\omega_0$, can induce pronounced valley-contrasting responses. Valleys are energy-degenerate band minima located at the different electron momentum K and K' (-K and +K) points of the Brillouin zone (Fig. \ref{fig:setup_compare}\textbf{g}). Because these valleys are widely separated in momentum space, intervalley transitions are difficult to achieve, allowing the valley index to serve as a robust discrete degree of freedom analogous to spin \cite{liu2019valleytronics}. This additional degree of freedom could be leveraged to extend information-processing capabilities beyond the conventional limits of Moore’s law. Valleytronic effects have been extensively studied using circularly polarized light \cite{mak2012control,tyulnev2024valleytronics,schaibley2016valleytronics}. They can also be induced by linearly polarized fields with broken time-reversal symmetry, such as few-cycle pulses \cite{jimenez2021sub} and two-color fields $\omega-2\omega$ \cite{hashmi2023enhancement}. Our results show that the BSV perturbation induces valley contrast in both ionization (i.e. excited population in the conduction band) and nonlinear optical emission.

Figure \ref{fig:valley} presents the electron responses along the K–$\Gamma$–K' direction ($K_y=0$). Figures \ref{fig:valley}\textbf{a} and \textbf{c} show the emitted harmonics versus $K_x$ for BSV center frequencies of $\omega_0$ and $0.5\omega_0$, respectively. The corresponding spectra integrated over the $K_x>0$ (contains K', blue) and $K_x<0$ (contains K red) are shown in Fig.\ref{fig:valley}\textbf{b} and Fig.\ref{fig:valley}\textbf{d}. The BSV centered at $\omega_0$ produces valley-symmetric emission, as in case "(i) None" (not shown), whereas the BSV centered at $0.5\omega_0$ induces a pronounced valley contrast.

Figures \ref{fig:valley}\textbf{e}–\textbf{g} show the $K_x$-resolved ionization dynamics driven by the classical pump alone and with BSV fields centered at $\omega_0$ and $0.5\omega_0$, respectively. The driving pulses are centered at $t=0$ (middle of the vertical axis). The ionization is initially zero and approaches a constant value after the pulse \cite{thorpe2023high}. Although the BSV centered at $\omega_0$ strongly modifies the ionization dynamics, the response remains valley symmetric (Figs. \ref{fig:valley}\textbf{e} and \textbf{f}). In contrast, the BSV centered at $0.5\omega_0$ both enhances ionization at certain electron momenta and induces a strong valley contrast, as suggested by Fig.\ref{fig:valley}\textbf{g}.

\subsection{Angular Dependence }
Another distinctive property of MoS$_2$ is its angular-dependent emission \cite{liu2017high,kobayashi2021polarization}. MoS$2$ possesses threefold ($C_3$) rotational symmetry about the $z$ axis, corresponding to rotations of $120^\circ$ in the $x$–$y$ plane. Here, we denote the polarization angle of the classical driving laser relative to the $x$-axis by $\phi$. Owing to the oscillation of a multi-cycle electric field, $\phi$ and $\phi+180^\circ$ are equivalent, reducing the symmetry to $60^\circ$ \cite{yue2022signatures,chang2024many,liu2017high}.

\begin{figure}[H]
\centering
    \includegraphics[width=0.7\linewidth]{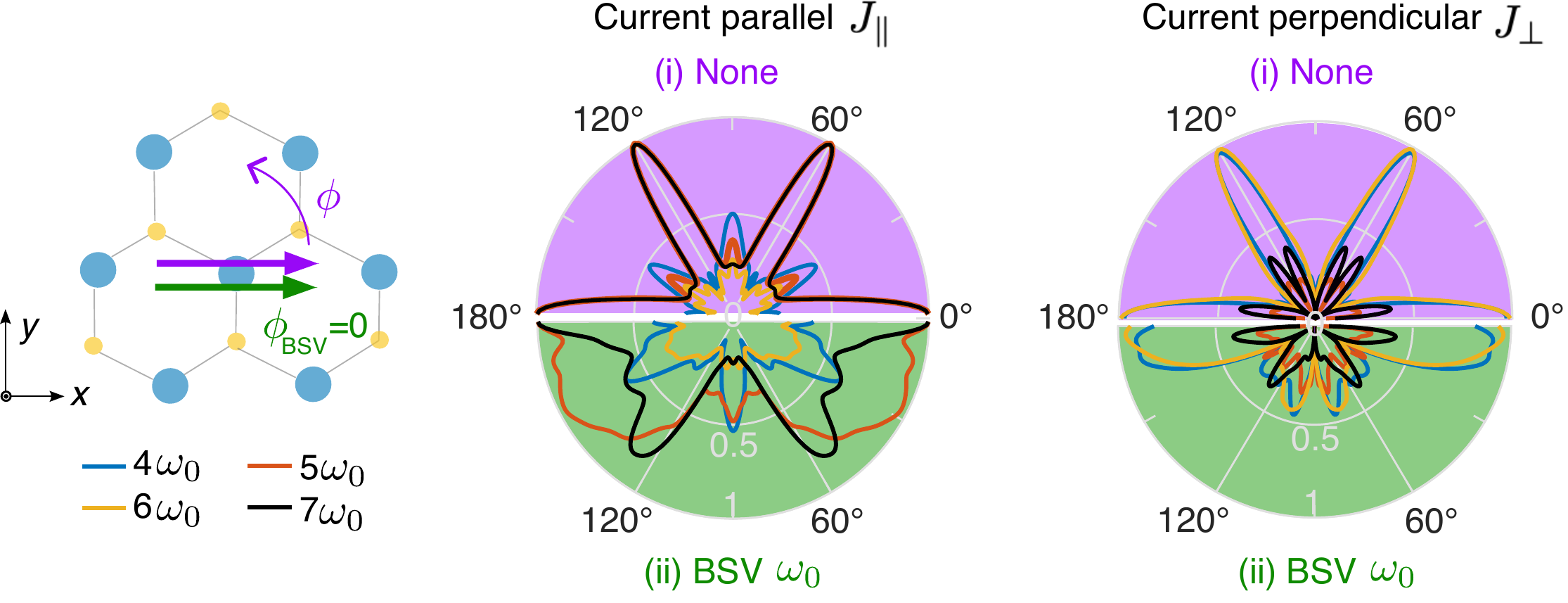}
\caption{Angular-dependent emission with and without BSV perturbation. Case "(i) None" with a classical field alone is presented in the upper hemisphere as a reference. Case "(ii) BSV" with center frequency $\omega_0$ is shown in the lower hemisphere. The BSV polarization is fixed along $x$, and only the classical field is rotated. The emission currents parallel $J_\parallel$ and perpendicular $J_\perp$ to the classical field are shown on the left-hand side and right-hand side, respectively. }\label{fig:angle}
\end{figure}

For analysis, we decompose the nonlinear current into components parallel ($J_{\parallel}$) and perpendicular ($J_{\perp}$) to the classical field polarization. For each harmonic order, both components are normalized by the maximum total current amplitude, $\sqrt{J_{\parallel}^2+J_{\perp}^2}$. Figure \ref{fig:angle} compares the classical field alone (upper semicircle) with the field perturbed by a BSV centered at $\omega_0$ (lower semicircle). The BSV remains polarized along $x$, while the classical-field polarization is rotated.

Figure \ref{fig:angle} suggests that odd harmonics dominate along the  $J_{\parallel}$ direction, whereas even harmonics dominate along the $J_\perp$ direction \cite{yue2022signatures,liu2017high,chang2024many}. Additionally, the BSV pulse can significantly alter the angular dependence of the harmonic emission, which exhibits $180^\circ$ symmetry. This new symmetry is unique due to the quantum perturbation. If a classical perturbation is used, one will observe an angular pattern similar to the upper hemisphere as shown in (i), but distorted towards the $x$ (bias induced by a classical perturbation), which does not have $180^\circ$ symmetry.

\section{Discussion and Conclusions}
We demonstrate that nonlinear emission and ionization in solids can be manipulated by a classical driving field accompanied by a very weak quantum field—bright squeezed vacuum (BSV). Remarkably, despite carrying less than $ 0.1\%$ of the classical field's energy, BSV induces distinct electron responses across the Brillouin zone, accessing regions otherwise inaccessible under purely classical excitation. In addition, BSV can introduce markedly different responses around the K and K' valleys, offering a route toward valley-selective control and information encoding—key ingredients for valleytronic technologies \cite{schaibley2016valleytronics}. Note that the effects induced via the BSV perturbation are unique and cannot be reproduced by a classical perturbation. Moreover, BSV introduces additional degrees of freedom—namely, its energy, center frequency, and polarization—that serve as tunable levers for controlling electron dynamics. Notably, we demonstrate that the BSV’s central frequency at $\omega_0$ and $0.5\omega_0$ induces distinctly different effects.

Finally, using 2D materials addresses growing demands for miniaturized photonic devices, in line with the prediction of Moore’s Law. Our work presents a versatile and scalable route toward chip-integrated quantum-optical diagnostics and control platforms. The low-energy nature of using BSV as a quantum dial offers further practical advantages. High-repetition-rate laser systems typically provide lower pulse energies, limiting access to extreme nonlinear regimes. By strongly extending the nonlinear response with only a weak BSV perturbation, our approach could make these regimes accessible using existing high-repetition-rate sources. The resulting reduction in the required pulse energy could lower experimental costs and mitigate material damage, while rapid signal averaging could improve the signal-to-noise ratio in time-resolved measurements.

\section{{Methods}}
\subsection{Calculations of the Nonlinear Current}
The nonlinear current is obtained by calculating the expectation value of the velocity operator as follows
\begin{equation}
J_i=-e\langle\Psi| \hat{v}_i|\Psi\rangle=-e\text{Tr}\left[{\rho}_{11} v_{i, 11}+{\rho}_{22} v_{i, 22}+\rho_{ 12} v_{i, 21}+{\rho}_{ 21} v_{i, 12} \right],
\end{equation}
where $i\in\{x,y\}$ denotes the current component along the $i$-direction,  $e=|e|$ is the elementary charge, and $\text{Tr}\left[\cdot\right]=\sum_l \bra{l}\, \cdot\ket{l}$ represents the trace of all the photon number states $\ket{l}$ which only operates on $\rho$. We restrict our analysis of the material to a two-band system, where the velocity tensor elements can be written as $v_{i,nm}\,$ with $n,m\in\{1,2\}$. Here, indices “2” and “1” label the conduction and valence bands, respectively. We solve the time evolution of the Hamiltonian analytically in the interaction frame. In particular, by using the Dyson expansion, we obtain a closed form of the density matrix up to the second order i.e. $\rho\approx\rho^{(0)}+\rho^{(1)}+\rho^{(2)}$. The nonlinear emission current is then decomposed into contributions corresponding to orders of the density matrix element as:
\begin{equation}
    j_i^{(n)}=-e\text{Tr}\left[
{\rho}^{(n)}_{11} v_{i, 11}+{\rho}^{(n)}_{22} v_{i, 22}+\rho_{ 12}^{(n)} v_{i, 21}+{\rho}_{ 21}^{(n)} v_{i, 12}  \right],\quad J_i=\sum_n j_i^{(n)}
\end{equation}
where $n\in\{0,\,1,\,2\}$ is the Dyson expansion order. Details of the analytical expressions of the density matrix elements can be found in Supplementary Material Section IV and our previous work \cite{boroumand2025strong}. After taking the trace and introducing the Rabi frequency $\Omega = (2e/\hbar) 
\mathbf{d}\cdot \mathbf{ E}$ where 
$\mathbf{d}$ is the transition dipole and $\mathbf{E}$ is the pump electric field, together with the band energy $\mathcal{E}$, we obtain:
\begin{align}
   & j_i^{(0)}=-e[v_{i,11}V_1^2+|V_2|^2v_{i,22}]-2e\text{Re}(V_1V_2v_{i,21})\label{j0}\\  
   & j_i^{(1)}=-2e(v_{i,22}-v_{i,11})\text{Re}\left[V_1V^*_2\int_{-\infty}^t  \exp{\left[2iS(t,t^\prime)\right]}\frac{i\Omega(t^\prime)}{2}\mathcal{R}(t,t^\prime)dt^\prime\right]\nonumber\\
&-2e\text{Re}\left[v_{i,21}V_1^2\int_{-\infty}^t  \exp{\left[2iS(t,t^\prime)\right]}\frac{i\Omega(t^\prime)}{2}\mathcal{R}(t,t^\prime)dt^\prime-v_{i,21}V_2^2\left(\int_{-\infty}^t  \exp{\left[2iS(t,t^\prime)\right]}\frac{i\Omega(t^\prime)}{2}\mathcal{R}(t,t^\prime)dt^\prime\right)^*\right]\label{j1}\\
& j^{(2)}_i=-\frac{e}{4}(v_{i,11}-v_{i,22})(V_1^2-|V_2|^2)\int_{-\infty}^t\int_{-\infty}^t  \exp{\left[2iS(t_1,t_2)\right]}\Omega^*(t_1)\Omega(t_2)\mathcal{R}(t_1,t_2)dt_1dt_2\nonumber \\
&-e\text{Re}\left\{v_{i,21}V_1V_2\int_{-\infty}^t\int_{-\infty}^t  \exp{\left[2iS(t_1,t_2)\right]}\Omega^*(t_1)\Omega(t_2)\mathcal{R}(t_1,t_2)dt_1dt_2\right\}.  \label{j2}
\end{align}
In Eqs.(\ref{j0}-\ref{j2}), $\text{Re}[\cdot]$ represents taking the real part. We solve Eqs. (\ref{j0})–(\ref{j2}) numerically, as it is, without invoking the saddle-point approximation. The calculation is parallelized over 48 CPU cores, reducing the computation time to approximately 6 hours.

Furthermore, both the band energy $\mathcal{E}$ and the Rabi frequency $\Omega$ are defined in the shifted Brillouin zone $\bm{K}_t=\bm{K}+e\bm{A}(t)/\hbar$, with crystal momentum $\bm{K}=(K_x,K_y)$, and the vector potential $\bm{A}(t)=-\int_{-\infty}^t \bm{E}(\tau)d\tau$ \cite{boroumand2025strong}. Besides, we have also used
\begin{align}
&S(t_1,t_2)=\int_{t_2}^{t_1} \mathcal{E}_s(\tau)/(2\hbar) d\tau,\quad \mathcal{E}_s= \sqrt{\mathcal{E}(t)^2+\hbar^2\Omega(t)^2}\\
&V_1=\sqrt{\frac{\mathcal{E}+\mathcal{E}_s}{2\mathcal{E}_s}},\quad V_2=\frac{-\hbar\Omega}{\sqrt{2\mathcal{E}_s(\mathcal{E}+\mathcal{E}_s)}}.
\end{align}
Details related to the response function $\mathcal{R}$ can be found in Section B below. One should also notice that
\begin{align}
& \text{FT}[J_i(t,K_x,K_y)]=J_i(\omega,K_x,K_y),\label{main:eq:Jtw}\\
  &    J_i(\omega)=\frac{1}{(2\pi)^2}\iint J_i(\omega,K_x,K_y)dK_xdK_y,\label{main:eq:intkxky}
\end{align}
where FT$[\cdot]$ represents the Fourier transform. The harmonic spectrum $|J_i(\omega)|$ is presented in Fig.\ref{fig:setup_compare}\textbf{a,b}. The momentum response distribution $J_x(7\omega_0,K_x,K_y)$ is shown in Fig.\ref{fig:setup_compare}\textbf{h-j}.

\subsection{Details Related to the Response Function $\mathcal{R}$}\label{main:method:response}
Our theoretical framework incorporates all effects induced by the external perturbation through the response function that emerges upon evaluating the trace involving $\rho$, which reads 
\begin{equation}
\mathcal{R}(t,t^\prime)=\text{Tr}\left[\mathcal{B}D^{\dagger}\left(t^\prime\right)^2 D^2(t)\right]=\exp{\left\{-\frac{1}{2}\left[ \bm{\beta}^\dagger(t,t^\prime) \cosh (2\bm{r})\bm{\beta}(t,t^\prime)+\text{Re}\left[\bm{\beta}^\dagger(t,t^\prime) \sinh( 2\bm{r}) \bm{\beta}^*(t,t^\prime)\right]
\right]\right\}},\label{main:eq:Rs}
\end{equation}
where $\mathcal{B}\propto \ket{\xi}\bra{\xi}$ denotes the density matrix of the external perturbation in the interaction frame. Here, $\ket{\xi}\bra{\xi}$ represents the BSV (see Supplementary Material, Sec.~II.A for details). Different perturbations can be described by choosing the corresponding $\mathcal{B}$. Furthermore, we have 
\begin{align}
  &  \bm{\beta}(t,t^\prime)=[\beta_1,\beta_2,\cdots],\quad \beta_q=\frac{2}{\hbar\omega_q}[g_q(\bm{K}_{t^\prime})\exp{(i\omega_qt^\prime)}-g_q(\bm{K}_t)\exp{(i\omega_qt)}]\\
  & D(t)=\exp{\left\{-\sum_{q}\left[\frac{g^*_{q}(\bm{K}_t)}{\hbar\omega_{q}}{b}_{q}^{\dagger}\exp{(i\omega_q t)}-\frac{g_{q}(\bm{K}_t)}{\hbar\omega_{q}}{b}_{q}\exp{(-i\omega_q t)}\right]\right\}},\\
  &g_q(\bm{K}_t)=\frac{e\left[v_{i,22}(\bm{K}_t)-v_{i,11}(\bm{K}_t)\right]}{4\pi c}\sqrt{\frac{\hbar\omega_q}{c\epsilon_0}d\omega_q},\label{main:eq:gq}
\end{align}
where $g_q$, $\hbar\omega_q$, $b_q$ are the coupling coefficient, photon energy, and boson annihilation operator of a given mode $q$, respectively, with $q\in\{1,2,3,\cdots N_q\}$. The angular-frequency spacing between adjacent modes in the plane-wave basis is denoted by $d \omega_q$.

Without loss of generality, we have chosen the BSV to have 31 modes i.e. $N_q=31$. It is important to notice that $\bm{r}$ and all its related functions such as $\cosh{2\bm{r}}$ and $\sinh{2\bm{r}}$ are rank-two tensors with size $N_q\times N_q$; $\beta$ is a column vector with size $N_q\times 1 $. Consequently, from Eq.(\ref{main:eq:Rs}), the response function $\mathcal{R}$ is a scalar function.

\section{Acknowledgments}
L.W. would like to show heartfelt gratitude to the Digital Research Alliance of Canada, which allows happy and intensive code running with almost no queuing time; to thank Milton and Rosalind Chang Pivoting Fellowship from the Optica Foundation; the Ministry of Education Singapore, Academic Research Fund Tier 2 (T2EP50125-0015); X-lites (NSF Award 2411691); Prof. Zenghu Chang, Dr. Andrei Rasputnyi and Dr. Igor Tyulnev for very helpful discussions. T.B. thanks NSERC and Air Force Office of Scientific Research (FA9550-18-1- 0223) for the support.
\bibliographystyle{apsrev4-1} 
\bibliography{apssamp} %
\end{document}